\documentstyle[12pt]{article}
\newcommand{\oae}{\overline{a}}
\newcommand{\oale}{\overline{\alpha}_s}

\def \asQ{\relax\ifmmode\bar{\alpha}_s(Q^2)
\else{$\bar{\alpha}_s(Q^2)${ }}\fi}
\def \alQ{\relax\ifmmode\bar  \alpha(Q^2)\else{$\bar \alpha(Q^2)${ }}\fi}
\def \as0{\relax\ifmmode\bar  \alpha_s(0)\else{$\bar   \alpha_s(0)${ }}\fi}
\def\albar{\relax\ifmmode{\bar{\alpha}}\else{$\bar{\alpha}${ }}\fi}
\def\albars{\relax\ifmmode{\bar{\alpha_s}}\else{$\bar{\alpha_s}${ }}\fi}
\def \alQa{\relax\ifmmode\bar\alpha(Q^2,\alpha)\else{$\bar\alpha(Q^2,
\alpha)${ }}\fi}
\newcommand{\beq}{\begin{equation}} \newcommand{\eeq}{\end{equation}}
\newcommand{\beqlab}{\begin{equation}\label}
\begin{document}
\begin{center}
\vspace{4.0cm}
{\Large\bf  Analytic QCD running coupling with finite   \\
              IR behaviour and universal \as0 value } \\
\vspace{8mm}
{\large D.V. Shirkov$^*$
and I.L. Solovtsov }\\
\vspace{0.2cm}
{\em  Bogoliubov Laboratory  of  Theoretical  Physics, \\
JINR, Dubna, 141980 Russia } \\
$^*$E-mail address: shirkovd@th-head.jinr.dubna.su   \\
\end{center}
\vspace{5mm}
\begin{abstract}
As is known from QED, a possible solution to the ghost-pole
trouble can be obtained by imposing the $Q^2$-analyticity imperative.
Here, the pole is compensated by the $\alpha$ non-analytic contribution
that results in finite coupling renormalization. \par
   We apply this idea to QCD and arrive at the $Q^2$ analytic
$\bar{\alpha}_s(Q)$. This solution corresponds to
perturbation expansion, obeys AF and, due to
non-perturbative contribution, has a regular IR behaviour. It does
not contain any adjustable parameter and has a finite IR limit
$\bar{\alpha}_s(0)$ which depends only on group symmetry factors.
In the one-loop approximation it is equal to $4\pi/\beta_0 \simeq 1.40\,$
and turns out to be surprisingly stable with respect to higher order
corrections. On the other hand, the  IR behaviour of our new analytic
solution agrees with recent global low  energy experimental estimates
of $\bar{\alpha}_s(Q^2)$.
\end{abstract}

\vspace{5mm}

{\bf 1.} We recall first some results obtained in QED about 40 years ago.
The QED effective coupling \alQ being proportional to the transverse
dressed photon propagator amplitude 
is an analytic function in the cut complex $Q^2$ plane and satisfies
the K\"allen-Lehmann spectral representation.
    The ``analytization procedure'' elaborated in papers \cite{red,trio}
consists of three steps:

(1) Find an explicit expression for $\albar_{RG}(Q^2)$ in the space-like
region by a standard RG improvement of perturbative input. Continue this
expression to the time-like $Q^2$ domain.

(2) Calculate the imaginary part of $\albar_{RG}(-Q^2)$ on the cut
and define the spectral density
$\rho_{RG} (\sigma ,\alpha)= {\rm Im}\; \albar_{RG}(-\sigma, \alpha)$.

(3) Using the spectral representation
 with $\rho_{RG}$ in the integrand define an analytic $\albar_{an}(Q)$.

For one-loop massless QED, this procedure produces~\cite{trio} an
explicit expression (see Eq.~(2.6) in Ref.~\cite{trio} or analogous
QCD Eq.~(\ref{an1}) below) which has the following properties:

(a) it has no ghost pole,

(b) considered as a function of $\alpha$ in the vicinity of the point
   $\alpha=0$ it has an essential singularity of the $\exp(-3\pi/\alpha)$
     type,

(c) in the vicinity of this singularity for real and positive $\alpha$
    it admits an asymptotic expansion that coincides with usual
    perturbation theory,

(d) it has a finite ultraviolet asymptotic limit,
    $\bar{\alpha}(\infty, \alpha) = 3\pi$, which {\it does not depend
    on the experimentally input value} $\alpha\simeq 1/137$.

    The same procedure in the two-loop massless QED approximation
yielded~\cite{trio} a more complicated expression with the same
essential features.
\vspace{1mm}

{\bf 2.}  To use the same technique in QCD one has to make two
observations.  First, since \asQ has to be defined via a product
of propagators and a vertex function, validity of the spectral
representation is not obvious. However, this validity has been
established in the Ref.\cite{ilya} on the basis of analytic
properties of the vertex diagrams. Second, for QCD with an arbitrary
covariant gauge running of the coupling and gauge parameter are
interconnected. For simplicity, we assume that the
$\overline{\rm MS}$ scheme is used.

In the following we  use the spectral representation in the
non-subtracted form
\begin{equation} \label{5}
\oae(Q^2)\,=\,\frac{\oale(Q^2)}{4\,\pi}\,=\,\frac{1}{\pi}\,\int_0^\infty\,
\frac{\rho (\sigma , \Lambda)\,d\sigma\,}{\sigma\,+\,Q^2\,-i\epsilon}\, .
\end{equation}

The usual massless one-loop RG approximation yields the spectral function
$$
\rho^{(1)}_{RG} (\sigma ,\Lambda)\,= \frac{\pi}{\beta_0\;({\ell}^2\,
+\,\pi^2)}\, , \qquad {\ell}\,=\,\ln\,\frac{\sigma}{\Lambda^2}\,
,~~~\beta_0=11-\frac{2}{3}n_f~.$$
Substituting $\rho^{(1)}_{RG}$ into spectral representation Eq.~(\ref{5})
we get
\begin{equation} \label{an1}
\oae ^{(1)}_{an,s}(Q^2)\,=\, \frac{1}{\beta_0}\, \left[\, \frac{1}
{\ln Q^2/\Lambda^2}\,+\,\frac{\Lambda^2}{\Lambda^2\,-\,Q^2}\,\right]\, ,
\end{equation}
where we have used the QCD scale parameter $\Lambda$.
The ``analytic'' running coupling  Eq.~(\ref{an1}) has
no ghost pole and its limiting  value
\beqlab{univ}
 \oale^{(1)}(0)\,=\; 4 \,\pi/\beta_0
 \eeq
does not depend on $\Lambda$ being an {\it universal constant which
depends only on group factors}. \par
We have become accustomed to the idea that theory supplies us with a
family of possible curves for \asQ and one has to choose the
``physical one" of them by comparing with experiment. Here, in
Eq.~(\ref{an1}) the whole bunch of possible curves for $\albars(Q^2)$
corresponding to different $\Lambda$ have the same limit at $Q^2=0$
as shown on Fig.~1. For $n_f=3$ it is equal to
\begin{equation} \label{as1}
\oale^{(1)}(0)=4\,\pi/9\simeq\,1.396 \,.
\end{equation}

Another feature of Eq.~(\ref{an1}) is the fact that its correct
analytic properties in the IR domain are provided by a
nonperturbative  contribution\footnote{The connection between
``$Q^2$ analyticity" and ``$\alpha$ nonperturbativity" has been
discussed in the Ref.~\cite{dv76}.} like $\exp (-1/a\beta_0)$.

To investigate the stability of the result, Eq.~(\ref{as1}), with
respect to the next loop corrections, we have considered the two-loop
approximation to \asQ in the form
$$
\bar{a}^{(2)}_{RG}=\frac{1}{\beta_0\; [L +b_1 \ln L]}~,~~
L=\ln \frac{Q^2}{\Lambda^2}~,~~b_1=\frac{\beta_1}{\beta_0^2}~\;,
~~\beta_1=102-\frac{38}{3}n_f $$
and
\begin{equation} \label{8}
\rho^{(2)}_{RG} (\sigma ,\Lambda)\,=\,\frac{I({\ell})}
{\beta_0 \;[R^2({\ell})\,+\,I^2({\ell})]}~ ,
~~{\ell}=\ln \frac{\sigma}{\Lambda^2}~,
\end{equation}
$$ R({\ell})\,= {\ell}\,+\, b_1\,\ln\,\left(\sqrt{{\ell}^2\,
+\,\pi^2}\right)\,~,~~
I({\ell})\,=\,\pi\,+\,b_1\,{\rm arccos}\,~ \frac{{\ell}}
{\sqrt{{\ell}^2\,+\,\pi^2}}\, .$$

The limiting two-loop value $\oale^{(2)}(0)$ is also specified only
by group factors via $\beta_0$ and $\beta_1 \,$.
Surprisingly, this value found by numerical calculation
practically coincides with the one-loop result.
For the $\overline{MS}$ scheme in a three-loop approximation
$\oale^{(3)}(0)$ changes by  a few percent. Thus, the value
$\oale(0)$ is remarkably stable with respect to higher loop
corrections and is practically independent of renormalization scheme.
\vspace{1mm}

{\bf 3.}
To fix $\Lambda$ we use the reference point $M_{\tau}=1.78\; {\rm GeV}$
with $\oale (M_{\tau}^2)=0.33\pm 0.03$~\cite{Narison}. Corresponding
solutions $\oale ^{(l=1,2,3)}(Q^2)$ are very close to each other
for the interval of interest $Q^2 \leq 10\,{\rm GeV^2}$.
For instance, at $Q^2=10\,\,{\rm GeV}^2 $ we have
$ \oale ^{(1)}(10) \simeq \oale^{(2)}(10)=0.267,~
~\oale ^{(3)}(10)=0.265~.$  Here, again we used $n_f=3$ as the average
number of active quarks in the spectral density. This seems to be
reasonable in the IR region.

For a more realistic description of the evolution of \asQ
in the Euclidean region $3\; {\rm GeV} < Q < 100\; {\rm GeV},$ one
should  take into account heavy quark thresholds.
  Using the explicitly mass-dependent RG formalism developed~\cite{56}
in the 50's a "smooth matching" algorithm has been devised
recently~\cite{mikh}. This can be given to correct  analytic properties
while incorporating heavy quark thresholds.

   The idea that \asQ  can be frozen at small momentum has been
recently discussed in some papers (see, e.g., Ref.~\cite{M-Stev}).
There seems to be experimental evidence indicating behaviour of this
type for the QCD coupling.
As the appropriate object for comparison with our theoretical
construction we use the  average
\begin{equation}
\label{Eq.12}
A(k)\,=\,\frac{1}{k}\,\int_0^k\,dQ\,\albars(Q^2, \Lambda)\, .
\end{equation}

"Experimental" estimates for this integral  are
$A(2\,\,{\rm GeV})=0.52 \pm 0.10$~GeV~\cite{Dok-Webb} and
$A(2\,\,{\rm GeV})=0.57 \pm 0.10$~GeV~\cite{dok}.
Our one-loop results in case~Eq.~(\ref{an1}) for $A$ are summarized
below for a few reference values of $\oale (M_{\tau}^2)$.

\vspace{1mm}
\begin{center}
\begin{tabular}{|l|c|c|c|}  \hline
$\oale (M_{\tau}^2)$&0.34&0.36&0.38 \\ \hline
$A(2\,\,{\rm GeV}) $&0.50&0.52&0.55 \\ \hline
\end{tabular}
\end{center}
\vspace{1mm}

Note here that a nonperturbative contribution, like the second term
in l.h.s. of Eq.~(\ref{an1}), reveals itself even at moderate $Q$ values by
"slowing down" the velocity of the \asQ evolution. For instance,
in the vicinity of $c$ and $b$ quark thresholds at $Q=3 ~ {\rm GeV}$
it contributes about 4\% which  could be essential for the resolution
of the "discrepancy"  between DIS and $Z_0$ data for \asQ.
\vspace{1mm}

{\bf 4.} We have argued  that a regular analytic behaviour for \asQ
in the IR region could be provided by nonperturbative contributions
which can be considered as a sum of powers of $\Lambda^2/Q^2$.

   Probably, our most curious result is the stability  of a
"long-range intensity of strong interaction", \as0, as well as the \asQ
IR behaviour that turns out to agree reasonably well with
experimental estimates.

 On the other hand, the nonzero \as0 value evidently contradicts the
 confinement property. To satisfy this, one should have $$ \as0=0 $$
 as it has recently been emphasized by Nishijima~\cite{nish} in the
 context of the connection between asymptotic freedom (AF) and color
 confinement (CC). \par

 It is possible to correlate this type of the IR limiting behaviour with
 the RG-improved perturbative input and $Q^2$-analyticity by inserting
 a Castillejo-Dalitz-Dyson zero into our solution
 (see Ref.~\cite{david}). Such a generalysed analytic
 solution will contain additional parameters. In this construction
 there is no evident relation between AF and CC. Here, CC is
 provided by nonperturbative contributions.

\vspace{1cm}
The authors would like to thank A.M.~Baldin, A.L.~Kataev
and V.A.~Rubakov useful comments.  Financial support of
I.S. by RFBR (grant 96-02-16126-a) is gratefully acknowledged.

 \vspace{2mm} 

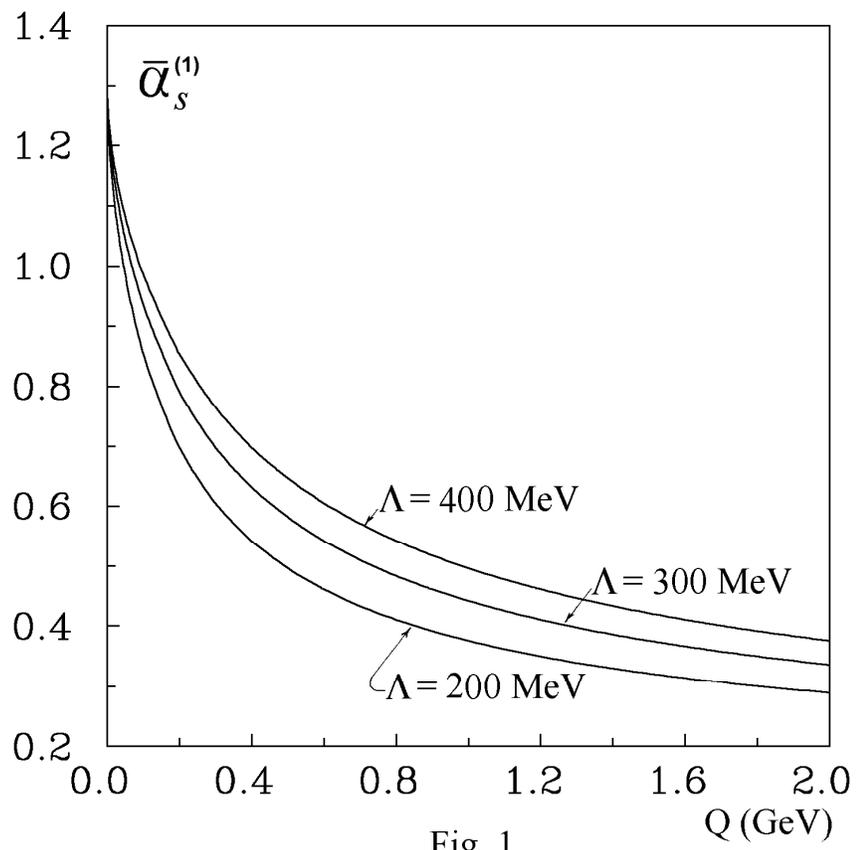
\end{document}